\documentstyle[aps,pre,eqsecnum,epsf]{revtex}
\newcommand{\be}{\begin{equation}}
\newcommand{\ee}{\end{equation}}
\newcommand{\bea}{\begin{eqnarray}}
\newcommand{\eea}{\end{eqnarray}}

\def \la{\langle}
\def \ra{\rangle}

\def \ts{ \textstyle }
\def \half{{\textstyle {1 \over 2}}}

\begin{document}

\title{Surface critical behavior of random systems at 
the special transition}

\author{Z. Usatenko$~{^{a,b,*}}$, Chin-Kun Hu$~{^{b,c,\$}}$}

\address{\it 
$~{^{a}}$ Institute for Condensed Matter Physics of the National 
Academy of Sciences of Ukraine Lviv, 79011, Ukraine \\ 
$~{^{b}}$ Institute of Physics Academia 
Sinica, Taipei, 11529, Taiwan \\ 
$~{^{c}}$ Department of Physics, National Dong Hwa 
University, Hualien 97401, Taiwan}
\vspace{-0.1cm}
\maketitle
\begin{abstract}

We study the surface critical behavior of 
semi-infinite quenched random Ising-like systems at the special transition 
using three dimensional massive field theory up to the two-loop 
approximation. Besides, we extend up to the next-to leading order, the 
previous first-order results of the $\sqrt{\epsilon}$ expansion obtained by 
Ohno and Okabe [Phys. Rev. B {\bf 46}, 5917 (1992)]. The numerical 
estimates for surface critical exponents in both cases are computed by 
means of the Pad\'e analysis. Moreover, in the case of the massive 
field theory we perform Pad\'e-Borel resummation of the resulting two-loop 
series expansions for surface critical exponents. The obtained results 
confirm that in a system with quenched bulk randomness the plane boundary 
is characterized by a new set of surface critical exponents. \end{abstract}

\renewcommand{\theequation}{\arabic{section}.\arabic{equation}}
\section{Introduction}
\setcounter{equation}{0}

The investigation of the critical behavior of the real physical systems is 
of considerable theoretical and experimental interest. As usual, the real 
physical systems are characterized by the presence of different kinds of 
imperfections. In common, the defects and impurities may be localized 
inside the bulk as well as at the boundary.

Historically the systematic investigation of the quenched disordered 
systems was initiated in the seminal works by Harris, Lubensky 
\cite{HL74,L75} and Khmelnitskii \cite{X75}. It is shown that 
investigation of the Ising-like systems is of main concern from the whole 
class of $O(n)$ symmetric $n$-vector models in $d$ dimensions, because they 
satisfy the Harris criterion \cite{Harris74}, which states that the 
presence of randomness is relevant for such pure systems which have a 
positive specific heat exponent $\alpha$. The introducing of the 
bulk dilution into a system shifts the critical temperature of the bulk 
phase transition and drives the system to another, `random' fixed point in 
which unconventional scaling behavior is observed. As it is confirmed by 
the Wilson's renormalization group and $\epsilon$ expansions 
\cite{HL74,L75,X75,GL76,Sh77,JK77}, the massive field theory in three 
dimensions \cite{Jug83,SSh81,Sh88}, experiments 
\cite{B...83,M...86,Bel00}, and Monte-Carlo simulations \cite{CS86,MLT86}, 
the critical behavior of three-dimensional disordered Ising-like systems is 
characterized by a new set of critical exponents \cite{mit1}. The case of 
the Ising model at $d=2$ is marginal, because in this case $\alpha=0$ and 
the correspondent logarithmic corrections to the power laws singularities 
of the pure model take place, as was confirmed in a series of papers 
\cite{Do95,Sh94,Se94,DoPP95,L90,ILSS98}.

The presence of a surface leads to the appearance of additional problems 
in critical phenomena. The most general review of critical behavior at 
surfaces and list of related publications are given in 
\cite{B83,D86,D97}. It is well known \cite{B83,D86,D97,DD81,OO85} that 
each surface universality class is defined by the bulk universality class 
and specific properties of a given boundary. At the present time 
three surface universality classes, called ordinary, special and 
extraordinary, are known \cite{D86,D97}. They correspond to the respective 
surface transitions which occur at the bulk critical point 
$m_{0}^2=m_{0c}^{2}$ \cite{mit2} and are characterized by different fixed 
points 
\be 
c_{0}^{*,ord}=+\infty, \quad\quad c_{0}^{*,sp}=c_{0}^{sp},\quad\quad 
c_{0}^{*,extr}=-\infty. 
\ee 
Here $c_{0}$ is so called `bare surface enhancement', which measures the 
enhancement of the interactions at the surface, and 
$(m_{0}^{2},c_{0})=(m_{0c}^{2},c_{0}^{sp})$ is a multicritical point, 
called special point. 

The influence of quenched surface disorder on the surface critical 
behavior was investigated in a series of theoretical works 
\cite{DN89,D98} and Monter Carlo calculations \cite{ILSS98,PS98}. General 
irrelevance-relevance criterion of the Harris type for short-range as 
well as for long-range correlated surface disorder was derived 
\cite{DN89}. In the case of special transition it has been demonstrated 
\cite{DN89,DSh98} that the fixed point describing the surface critical 
behavior of three dimensional pure systems is stable with respect to 
short-range correlated surface disorder. Thus, the weak short-range 
surface disorder is irrelevant for three dimensional systems, 
but long-range correlated random field disorder is relevant in $d \le 4$ 
dimensions.
 
In a recent paper \cite{UShH00}, we quantitatively confirm the previous 
results by Ohno and Okabe \cite{OO92} that the introducing of 
quenched bulk randomness in semi-infinite systems bounded by a plane 
surface affects the surface critical behavior of these systems. 
 Thus obtained surface critical exponents of 
 quenched dilute semi-infinite systems at the ordinary transition 
 differ from the surface critical exponents of the pure 
 semi-infinite systems \cite{DSh98}. Besides, we showed that to order 
 $\epsilon$, the $\sqrt{\epsilon}$ expansion for surface critical exponents 
 $\eta_{\parallel}$ and $\eta_{\perp}$ has given negative value 
 of the correlation function critical exponent $\eta$ for the random bulk 
 Ising system according to the scaling relation $\eta = 
 2\eta_{\perp}-\eta_{\parallel}$. It confirms the well known fact that the 
 second order of the $\sqrt{\epsilon}$ expansion is not 
 enough to give correct positive value of critical exponent $\eta$ 
 \cite{Sh77,JK77,B87,McK94,ShAnSo97,FHY99,FHY00}. The obtained results 
 \cite{UShH00} have shown that this kind of deficiencies do not appear at 
 the calculations using the massive field-theoretic approach directly in 
 $d=3$ dimensions \cite{Par80}.

 All these have stimulated us to perform the investigation of the special 
transition occurring in quenched bulk dilute semi-infinite systems bounded 
with a plane surface. It should be mentioned that the problem of 
investigation of the special transition is very important from such point 
of view that at some conditions it may be reduced to the problem of the 
adsorption of $\theta$ polymers on a wall \cite{ED87,HG94}.  

The investigation of the critical behavior of the systems with quenched, 
i.e., time independent, randomness is possible to split on two directions. 
One of 
 them is renormalization-group approach introduced by Harris and Lubensky 
 \cite{HL74}. This approach involves applying the renormalization-group 
 transformation to the random system directly and subsequent averaging 
 over disorder. Ohno and Okabe \cite{OO92} employed the above mentioned 
 method to analyze the influence of randomness on the surface critical 
 behavior at $d=4-\epsilon$ dimensions in the frames of $\sqrt{\epsilon}$ 
 expansion. 
 
 Another technique introduced by Grinstein and Luther \cite{GL76} involves 
 firstly removing the randomness by averaging, and subsequent employing the 
 renormalization group. They considered an $mn$ vector model and showed 
 that analytic continuation of this model to $n=0$ is equivalent to a model 
 of  a random $m$-component spin system. An elegant derivation of this 
 equivalence has been given by Emery \cite{E75}. Our main calculations 
 are performed with this technique to treat randomness.

The present paper is dedicated to the investigation of the special surface 
transition in semi-infinite, quenched dilute Ising-like systems at the 
bulk `random' critical point directly in $d=3$ dimensions using the 
massive field theory up to the two-loop approximation. 
Besides, we extend up to the next-to leading order of the $\sqrt{\epsilon}$ 
expansion, the previous first-order results obtained by Ohno and Okabe 
\cite{OO92}. The numerical estimates for surface critical exponents of the 
special transition in both cases are calculated using extensive Pade 
analyses. Moreover, in the case of the massive field theory we  
performed Pad\'e - Borel resummations of the resulting two-loop series 
expansions and obtained quite reasonable and reliable numerical estimates 
for surface critical exponents. The obtained results confirm that in the 
case of quenched bulk randomness in semi-infinite systems the new set of 
the surface critical exponent appears.

\renewcommand{\theequation}{\arabic{section}.\arabic{equation}}
\section{Model}
\setcounter{equation}{0}

In the previous work \cite{UShH00}, we presented 
some notations about possibilities to use effective 
Landau-Ginzburg-Wilson Hamiltonian with cubic anisotropy defined in 
semi-infinite space 

\bea
H_{LGW}[\vec\phi]&=&\!\!\int_0^\infty\!\!\!\! dz\!\!\int\!\! d^{d-1}r 
\Big [ \half |\nabla\vec\phi |^2 +\half m_0^2 |\vec\phi 
|^2\nonumber\\
 &+&\ts{1\over 4!}v_0\sum\limits_{\alpha=1}^n \phi_{\alpha}^4 +
\ts{1\over 4!}u_0(|\vec\phi|^2)^2 \Big ],\label{1}
\eea
for description of surface critical behavior of quenched dilute 
semi-infinite Ising-like systems at the ordinary transition in the 
replica limit $n \to 0$. It should be mentioned that here  
$\vec{\phi}$ is an $n$ - vector field with the components $\phi_{i}, 
i=1,...,n$ defined on a half-space $I\!\!R^d_+\equiv\{{\bf 
x}{=}({\bf r},z)\in I\!\!R^d\mid {\bf r}\in I\!\!R^{d-1}, z\ge 0\}$ 
bounded by a plane free surface at $z=0$.

For the first time the above model was introduced by Grinstein and Luther 
\cite{GL76}. The $O(n)$ symmetric term in (\ref{1}) arises in the process 
of the configurational averaging of the free energy over disorder using the 
replica trick 

\be
F = -T \lim_{n\to 0} {1\over n}(\la Z^n \ra_{conf}-1 ),\label{2}
\ee
via cumulant expansion and in accordance with this its coupling constant 
$u_0\propto -\Delta < 0$. Here $Z$ is the partition function with Boltzman 
weight $e^{-H[\phi]}$, where $H[\phi]$ is the effective 
Landau-Ginzburg-Wilson Hamiltonian of an original Ising system with scalar 
field $\phi=\phi(x)$ 
\be
H[\phi]=\!\!\int_0^\infty\!\!\!\! dz\!\!\int\!\! d^{d-1}r \left [
\half |\nabla\phi |^2 +\half \tau_0 \phi^2 + \ts{1\over 
4!}v_0\phi^4 \right ].\label{3}
\ee
Parameter $\tau_0$ involves local random temperature fluctuations 
$\delta \tau(x)$ via $\tau_0=m_0^2+\delta \tau(x)$, where 
$\langle\delta\tau(x)\rangle_{conf}=0$ and 
$\langle\delta\tau(x)\delta\tau(x')\rangle_{conf}=\Delta\delta(x-x')$ 
with $\Delta >0$. It should be noticed that $m_0^2$ corresponds to a "bare 
mass" representing linear temperature deviations from the mean - field 
critical temperature.

In the present paper we wish to investigate the 
surface critical behavior of semi-infinite Ising-like systems with 
quenched bulk randomness at the special transition. The critical behavior 
at the special transition has own peculiarities.  
In general case effective Hamiltonian for such systems must involve terms 
describing surface interactions \cite{B83,D86,DD81,DDE83,DN89} 
\be 
U_{1}(\phi) = \frac{1}{2}c_{0}\vec{\phi}^2 - \vec{h} \vec{\phi}.\label{5} 
\ee
Usually these external surface fields $\vec{h}$ are position and time 
dependent, so that means they may be used as sources. But in the present 
investigation we restrict ourselves to the case $\vec{h}= 0$. Thus, the 
common form of the effective Hamiltonian describing the surface critical 
behavior of quenched dilute semi-infinite Ising-like systems in the 
replica limit $n \to 0$ can be written in the form 

\begin{eqnarray} 
H(\vec{\phi}) & = &\int_{0}^\infty dz \int d^{d-1}r [\frac{1}{2} 
\mid \nabla\vec{\phi} \mid ^{2} +  
\frac{1}{2} m_{0}^{2}\mid \vec{\phi} \mid^{2}\nonumber\\
&+& \frac{1}{4!} v_{0} \sum_{i=1}^{n} \phi_{i}^{4} + \frac{1}{4!}u_{0} 
(\mid \vec{\phi}\mid^{2})^{2})] + \frac{1}{2} \int d^{d-1}r  
c_{0}\vec{\phi}^{2}.\label{6} 
\end{eqnarray}  

It should be mentioned that fields $\phi({\bf r},z)$ 
satisfy the Neumann boundary condition \cite{DD81,DDE83}, so we have at 
$z=0$ that $\partial_{z}\phi({\bf r}, z)=0$. This Hamiltonian takes into 
account surface interaction in the form of additional term $\frac{1}{2}\int 
d^{d-1}r c_{0}\vec{\phi}^2$. The model defined in (\ref{6}) is restricted 
to translations parallel to the boundaring surface. Thus, only parallel 
Fourier transformations in $d-1$ dimensions take place.

\renewcommand{\theequation}{\arabic{section}.\arabic{equation}}
\section{Renormalization of the correlation function}
\setcounter{equation}{0}

The correlation function of the model (see Eq.(\ref{6})), which  
involves $N$ fields $\phi({\bf{x}}_{i})$ at distinct points 
${\bf{x}}_{i}(1\leq i \leq N)$ in the bulk and $M$ field 
$\phi({\bf{r}}_{j},z=0)\equiv \phi_{s}({\bf{r}}_{j})$ at distinct 
surface points with parallel coordinates ${\bf{r}}_{j}(1\leq j \leq M)$, 
has the form

\be
G^{(N,M)}(\{{\bf x}_{i}\}\{{\bf{r_{j}}}\}) = \la \prod_{i=1}^{N}
\phi({\bf x}_{i})\prod_{j=1}^{M}\phi_{s}({\bf r}_{j})\ra. \label{7}
\ee

The corresponding full free propagator in the ${\bf{p}} z$ representation 
is given by 

\be
G({{\bf{p}}},z,z') = \frac{1}{2\kappa_{0}} \left[ e^{-\kappa_{0}|z-z'|} - 
\frac{c_{0}-\kappa_{0}}{c_{0}+\kappa_{0}} e^{-\kappa_{0}(z+z')} 
\right],\label{8} 
\ee
with the standard notation 
$
\kappa_{0}=\sqrt{p^{2}+m_{0}^{2}}.
$
Here, ${\bf p}$ is the value of parallel momentum associated with $d-1$ 
translationally invariant directions in the system.
The first term in (\ref{8}) corresponds to usual free bulk propagator in 
coordinate space, between the points ${\bf r}=({\bf p},z)$ and ${\bf 
r}^{'}=({\bf p},z^{'})$, and the second, so called `surface' term, depends 
on the distance between the point ${\bf r}$ and its `mirror image' ${\bf 
r}^{'}=(0,-z^{'})$.  

The formulation of the randomness problem in the spirit of approach 
introduced by Grinstein and Luther indicate that the renormalization 
process for the random systems is similar to the `pure' 
case \cite{D86,DSh98}. As it is known \cite{D86,DSh98}, in the theory of 
semi-infinite systems the bulk field $\phi({\bf x})$ and the surface 
field $\phi_{s}({\bf r})$ should be reparameterized by different uv-finite 
renormalization factors $Z_{\phi}(u,v)$ and $Z_{1}(u,v)$ 
$$ \phi(x) = 
Z_{\phi}^{1/2}\phi_{R}(x)\quad\quad\quad and \quad\quad\quad 
 \phi_{s}(r)=Z_{\phi}^{1/2}Z_{1}^{1/2}\phi_{s,R}(r). $$

The renormalized correlation function involving N bulk and M 
surface fields $(N,M)\ne (0,2)$ can be written as 

\be
G_{R}^{(N,M)} (0 ; m,u,v,c)=Z_{\phi}^{-(N+M)/2} Z_{1}^{-M/2} 
G^{(N,M)} (0 ; m_{0},u_{0},v_{0},c_{0}).\label{9}
\ee

In order to obtain the critical exponent $\eta_{\parallel}^{sp}$ which 
characterizes surface correlations at special transition, it is sufficient 
to consider a two-point correlation function of surface fields 
$G^{(0,2)}(p)=\la\varphi(p,z=0)\varphi(-p,z^{'}=0)\ra$.

It should be mentioned that the renormalized mass $m$, coupling constants 
$u$, $v$, and the renormalization factor $Z_{\phi}$ are fixed via the  
standard normalization conditions of the infinite-volume theory 
\cite{Brezin76,GL76,Par80,PV00}.
 In order to remove short-distance singularities of 
the correlation function $G^{(0,2)}$ located in the vicinity of the 
surface and to define the surface-enhancement shift $\delta c$ and surface 
renormalization factor $Z_{1}$, new normalization conditions should be 
introduced. We normalize the renormalized surface two-point correlation 
function in such a manner that at zero external momentum it should 
coincide with the lowest order of perturbation expansion of the surface 
susceptibility $\chi_{\parallel}(p)=G^{(0,2)}(p)$
\be
G^{(0,2)} (p;m_{0},u_{0},v_{0},c_{0}) = \frac{1}{c_{0}+\sqrt{p^{2} + 
m_{0}^{2}}} + O(u_{0},v_{0})\label{12b}
\ee
and its first derivatives with respect to $p^2$. 
Thus we obtain necessary surface normalization conditions

\be
G_{R}^{(0,2)}(0;m,u,v,c) = \frac{1}{m+c}\label{10}
\ee
and 

\be
\left.\frac{\partial G_{R}^{(0,2)} (p;m,u,v,c)}{\partial p^{2}} 
\right|_{p=0} = - 
\frac{1}{2m(m+c)^{2}}. \label{11}
\ee

Eq.(\ref{10}) defines the required surface-enhancement shift $\delta c$. 
From Eq.(\ref{10}) is easy to see that the surface susceptibility 
diverge at $m=c=0$. This point corresponds to the multicritical 
point $(m_{0c}^{2},c_{0}^{sp})$ at which special transition takes place.

Taking into account the above normalization condition (\ref{11}) and 
expression for renormalized correlation function (\ref{9}) it is possible 
to define the renormalization factor $Z_{\parallel} = Z_{1} Z_{\phi}$ via  

\be
Z_{\parallel} = \left. 2m \frac{\partial}{\partial 
p^{2}}[G^{(0,2)} (p)]^{-1}\right|_ {p^2=0} = \lim_{p\to 
0}{m\over p}{\partial\over\partial p} [G^{(0,2)}(p)]^{-1}.\label{12} 
\ee

It should be mentioned that the next singular behavior of 
these renormalization $Z$ factors in the critical region takes place 
\begin{eqnarray}
&& Z_{\phi} \propto m^{\eta},\nonumber\\
&& Z_{1}^{sp} \propto m^{\eta_{1}^{sp}},\label{13}
\end{eqnarray}
where $m$ is identified via the inverse bulk correlation length 
$\xi^{-1}\propto t^{\nu}, \quad t=\frac{(T-T_{c})}{T_{c}}$. 
Here $\eta$ is the standard bulk correlation exponent and 
exponent $\eta_{1}^{sp}$ is specific for our quenched random semi-infinite 
system. As it is known \cite{Sh97,DSh98} these exponents $\eta$
and $\eta_{1}^{sp}$ arise as a RG arguments of an inhomogeneous 
Callan-Symanzik equation for correlation functions $G_{R}^{(0,2)}$ 
(\ref{9}) 
\be
\eta_{\phi}=m\!{\partial\over \partial m} 
\left.\ln\!{Z_{\phi}}\,\right|_{F\!P},\quad\quad\quad 
\eta_{1}^{sp}=m\!{\partial\over \partial m} 
\left.\ln\!{Z_{1}}\,\right|_{F\!P}.\label{14}
\ee
The simple scaling dimensional analysis of $G_{R}^{(0,2)}$ and mass 
dependence of $Z$ factors (\ref{13}) defines the surface correlation 
exponent $\eta_{\parallel}^{sp}$ via 
\be
\eta_{\parallel}^{sp}=\eta_{1}^{sp}+\eta. \label{15}
\ee
According to (\ref{12}),(\ref{14}) and (\ref{15}) we obtain for surface 
correlation exponent $\eta_{\parallel}^{sp}$

\begin{eqnarray}
\eta_{\parallel}^{sp}&=&m\!{\partial\over \partial m} 
\left.\ln\!{Z_{\parallel}}\,\right|_{F\!P}\nonumber\\
&=&\beta_u(u,v){\partial\ln Z_{\parallel}(u,v)\over \partial u}+
\left.\beta_v(u,v){\partial\ln Z_{\parallel}(u,v)\over \partial 
v}\,\right|_{F\!P}. \label{16} 
\end{eqnarray}

The above value should be calculated at the infrared-stable random fixed 
point (FP) of the underlying bulk theory. The other surface critical 
exponents of the special transition can be determined via the set of 
surface scaling relations \cite{D86}.

\renewcommand{\theequation}{\arabic{section}.\arabic{equation}}
\section{The perturbation series up to two-loops.} 
\setcounter{equation}{0}

As was indicated in the previous section, the calculation of the surface 
critical exponent $\eta_{\parallel}^{sp}$ can be performed in accordance 
with Eq.(\ref{16}), where renormalization factor $Z_{\parallel}$ is 
defined via (\ref{12}). It should be mentioned that 
here, by analogy with the infinite-volume theory, we start from inverse 
surface correlation function $[G_{R}^{(0,2)}]^{-1}$ in order to avoid  
the external lines depending both on the external momentum $p$ and the 
surface enhancement $c_{0}$ in each external propagator
\be
G(p;z,0) = \frac{e^{-\kappa_{0} z}}{\kappa_{0}+c_{0}}.\label{17}
\ee
In other words such procedure of the correlation function inversion allows 
to amputate the denominators $\kappa_{0} + c_{0}$ from each external 
propagator. Thus we consider the Feynman diagram expansion of the 
unrenormalized surface correlation function $[G^{(0,2)}]^{-1}$ in terms of 
the free propagator (\ref{8}) up to the two-loop order 

\begin{eqnarray}
&&[G^{(0,2)}(p;m_0,u_0,v_0)]^{-1}=c_{0}+\kappa_{0} - \epsfxsize=0.9cm 
\epsfbox{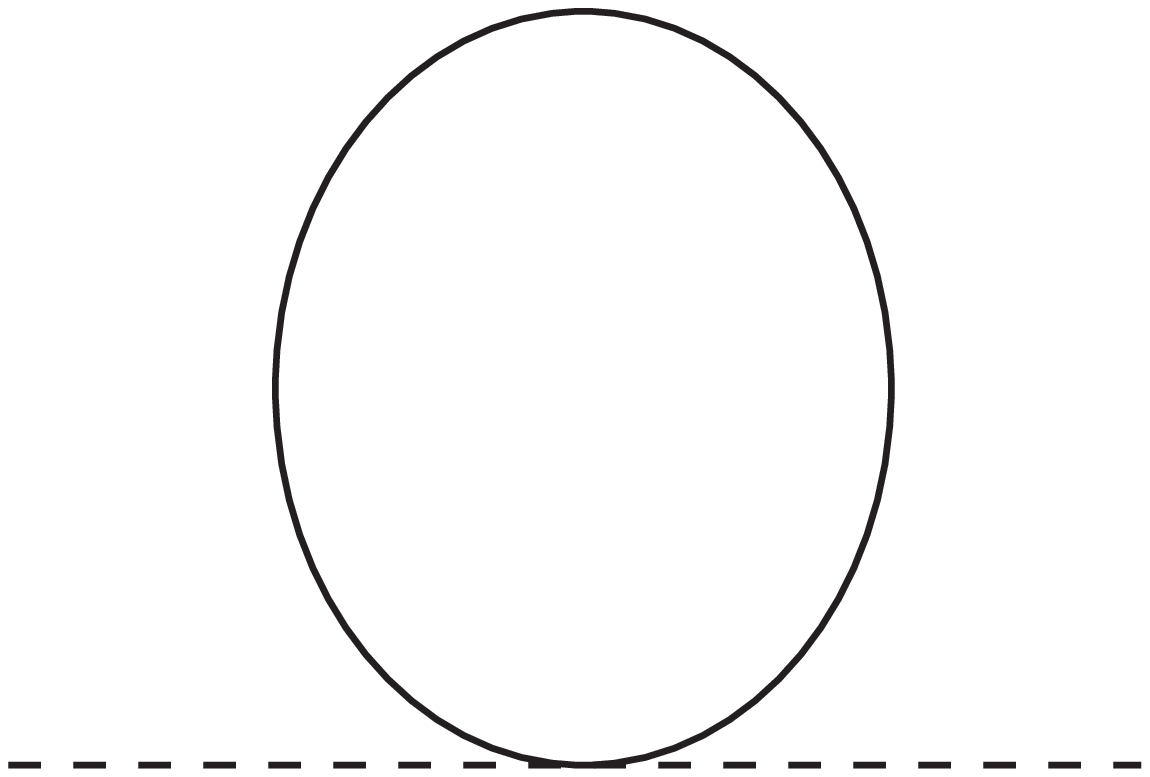}-[c_{0}+\kappa_{0}]^{-1}[\epsfxsize=0.9cm 
\epsfbox{fig1.eps}]^{2}\nonumber\\ 
&&-\raisebox{-5pt}{\epsfxsize=1.4cm \epsfbox{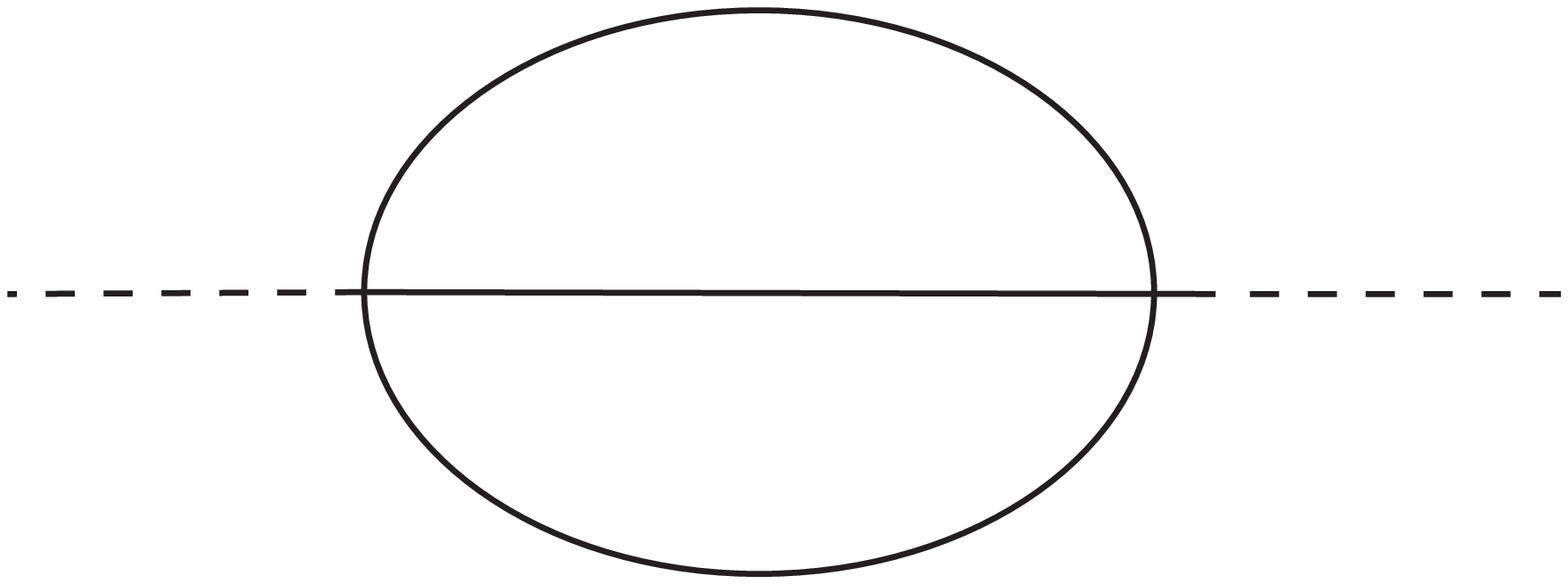}}- \epsfxsize=1.4cm 
\epsfbox{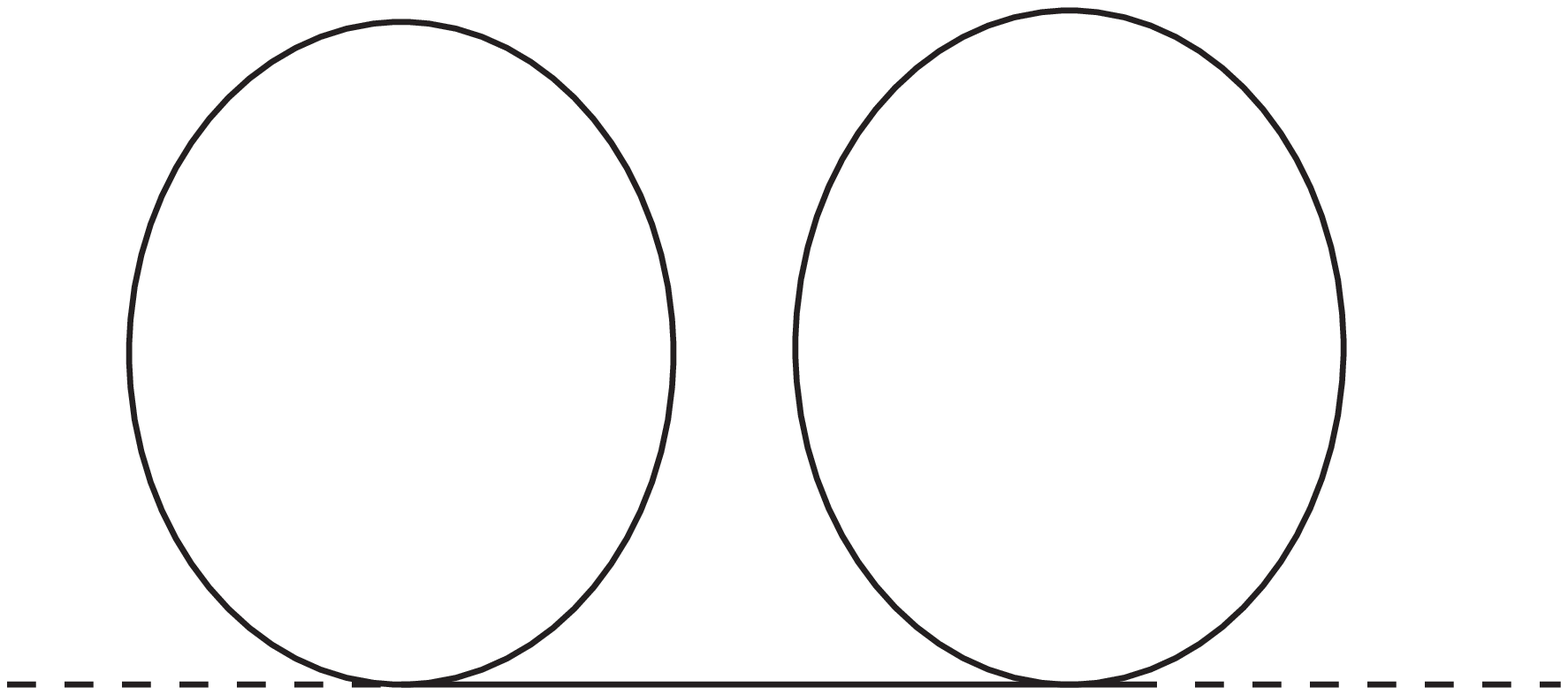}- \epsfxsize=0.9cm \epsfbox{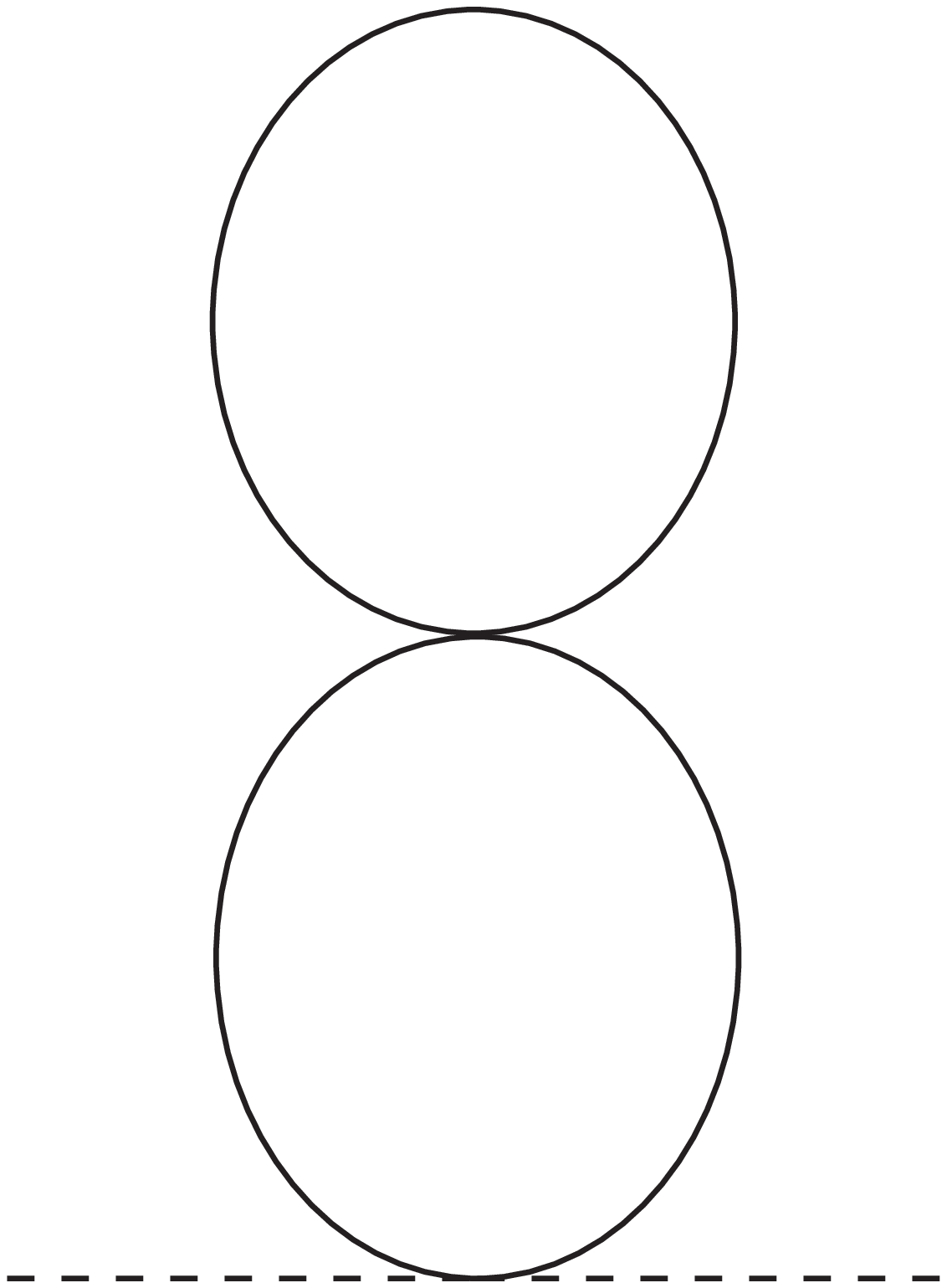}.\label{18}
\end{eqnarray}
Here the full internal lines denote the full free propagator (\ref{8}) 
and the dashed external lines correspond to the factors $e^{-\kappa_{0} 
z}$. In accordance with the presence solely $d-1$ translational invariance 
in (\ref{6}) only a partial cancellation in the external legs takes place 
and as a result the one-line reducible graphs do not disappear. 
All these graphs in (\ref{18}) have their own weights arising from the 
standard symmetry factors of the effective Hamiltonian (\ref{6})

\begin{mathletters}\label{ti}
\begin{eqnarray}
&&{- t_1^{(0)}\over 2}\quad\mbox{with}\quad t_1^{(0)}=\frac{n+2}{3}\,u_0+ 
v_{0}\,,\\ &&{t_2^{(0)}\over 6}\quad\mbox{with}\quad 
t_2^{(0)}=\frac{n+2}{3}\, u_{0}^{2}+ v_{0}^{2}+ 2v_{0}u_{0}\,,\\
&&{t_3^{(0)}\over 4}\quad\mbox{and}\quad {t_4^{(0)}\over 
4}\quad\mbox{with}\quad t_3^{(0)}=t_4^{(0)}=\big(t_1^{(0)}\big)^2,
\end{eqnarray}
\end{mathletters}
where $-t_{1}^{(0)}$ corresponds to the one-loop diagram, $t_{2}^{(0)}$ to 
the two-loop melon-like diagram, $t_{3}^{(0)}$ and $t_{4}^{(0)}$ 
to the reducible and irreducible two-loop diagrams in 
(\ref{18}), respectively. With a view to avoid the usual bulk uv 
singularities, which are present in (\ref{18}), we perform the mass 
renormalization (see \cite{UShH00}). After the mass 
renormalization we obtain that the divergent parts of the correlation 
function (\ref{18}) associated with the bulk terms of the free propagator 
(\ref{8}) cancel and the "melon-like" diagram becomes subtracted. Thus 
we obtain the next expression for inverse correlation function 
$[G^{(0,2)}]^{-1}$ 

\begin{eqnarray}
&&[G^{(0,2)}(p;m,u_0,v_0)]^{-1}=c_{0}+\kappa - \epsfxsize=1.1cm 
\epsfbox{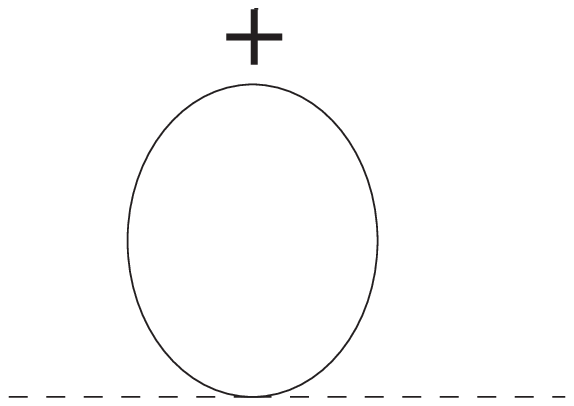}\nonumber\\ 
&&-\raisebox{-5pt}{\epsfxsize=1.7cm \epsfbox{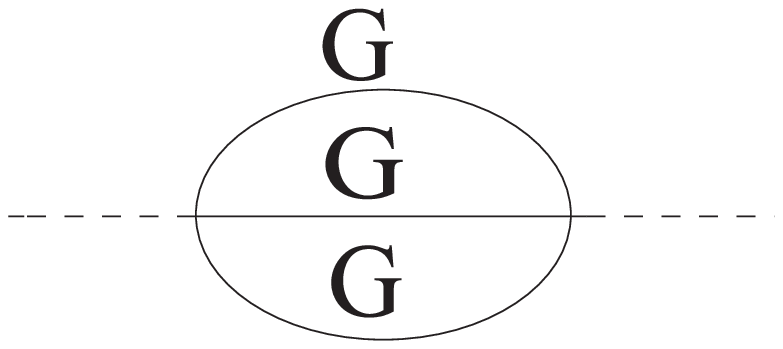}}
+\frac{1}{2\kappa}\raisebox{-5pt}{\epsfxsize=0.9cm \epsfbox{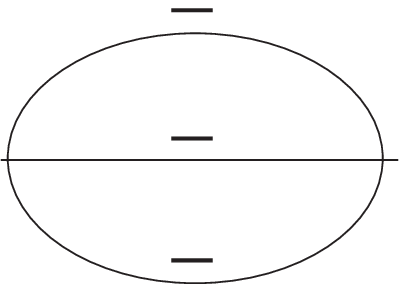}}-
\frac{m^2}{2\kappa}\frac{\partial}{\partial k^2}\left. 
\raisebox{-5pt}{\epsfxsize=1cm 
\epsfbox{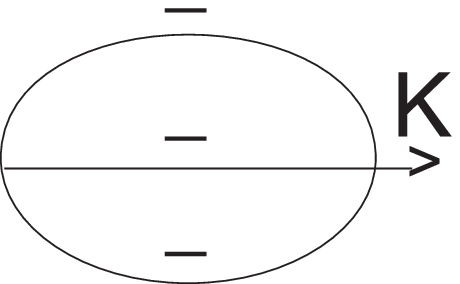}}\right|_{k^2=0}\nonumber\\ 
&&-\epsfxsize=2.1cm 
\epsfbox{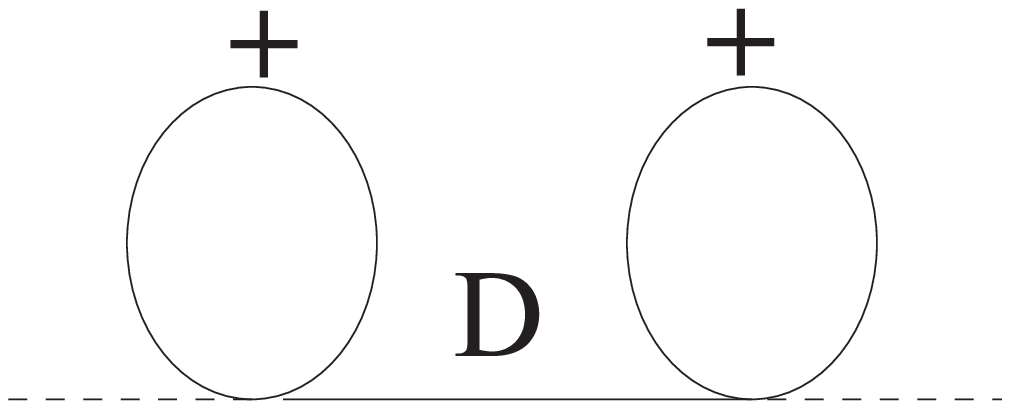}-\epsfxsize=1.1cm \epsfbox{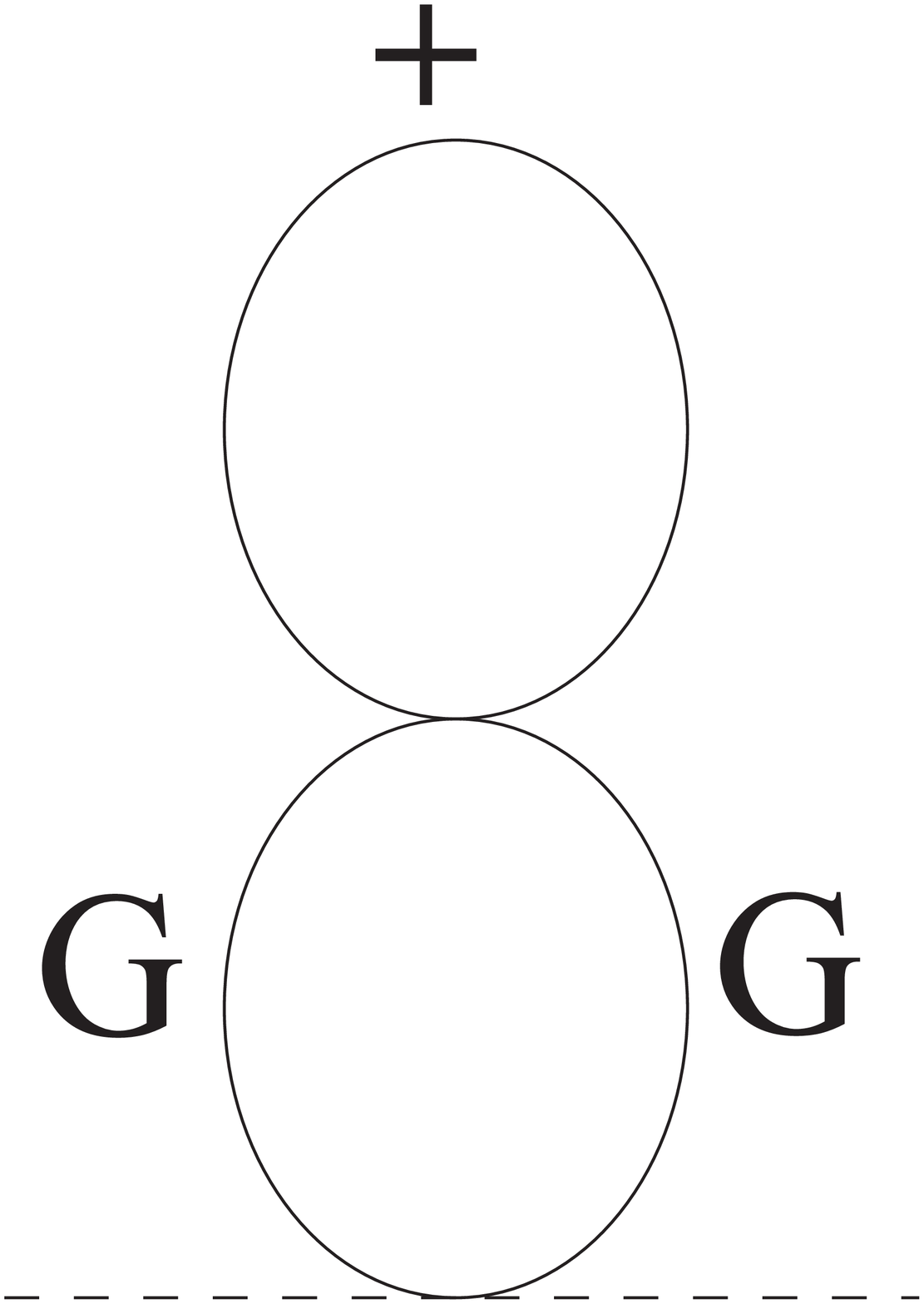}.\label{19} 
\end{eqnarray} 

The superscript $G$ means the full free propagator (\ref{8}), its 
bulk and surface parts are denoted with "-" and "+" signs, respectively. 
The line labeled by $D$ represents the Dirichlet propagator 

\be
G({{\bf{p}}},z,z') = \frac{1}{2\kappa_{0}} \left[ e^{-\kappa_{0}|z-z'|} - 
e^{-\kappa_{0}(z+z')} \right],\label{20} 
\ee
which arises from the difference 
\be
\epsfxsize=1.5cm 
\epsfbox{fig3m.eps}-[c_{0}+\kappa_{0}]^{-1}[\epsfxsize=0.9cm 
\epsfbox{fig1.eps}]^{2} = \epsfxsize=2cm 
\epsfbox{fig33s.eps}.\label{21}
\ee

The expansion (\ref{19}) still holds the uv divergences which are connected 
with the presence of the surface in the system. These divergent parts 
represented by the closed lines with the index "+".  The short-distance 
singularities of the inverse correlation function $[G^{(0,2)}(p)]^{-1}$ 
can be removed via the renormalization of the surface enhancement. The 
surface enhancement renormalization is defined via the normalization 
condition (\ref{10}). This condition can be rewritten for our inverse 
unrenormalized  correlation function in the form 
\be
Z_{\parallel}[G^{(0,2)}(0;m_0,u_0,v_0,c_0)]^{-1}=m + c.\label{22}
\ee
Substituting $[G^{(0,2)}(p)]^{-1}$ (see Eq.(\ref{19})) at zero 
external momentum $p=0$ into the normalization condition (\ref{22}) gives 
us the equation for the surface enhancement shift $\delta c$ 
\be
\delta c = (Z_{\parallel}^{-1}-1)(m+c) + \sigma_{0}(0;m,c_0=c+\delta 
c).\label{23}
\ee
This equation can be resolved using method of sequential  
iteration to determine the dependence of $c_{0}=c+\delta c$ on $c$ and 
$m$. Here $\sigma_{0}(0;m,c_0)$ denotes the sum of all loop diagrams in 
(\ref{19}). Among them $\sigma_{1}$ corresponds to the one-loop graph, 
$\sigma_{2}$ denotes the melon-like two-loop diagrams
\be
\sigma_{2}(p;m,c_{0}) = \raisebox{-5pt}{\epsfxsize=1.7cm \epsfbox{fig22s.eps}}
-\frac{1}{2\kappa}\raisebox{-5pt}{\epsfxsize=0.9cm \epsfbox{fig2b.eps}} +
\frac{m^2}{2\kappa}\frac{\partial}{\partial k^2}\left. 
\raisebox{-5pt}{\epsfxsize=1cm 
\epsfbox{fig2k.eps}}\right|_{k^2=0},\label{23a}
\ee

$\sigma_{3}$ and $\sigma_{4}$ represent the last two terms in 
Eq.(\ref{19}), respectively.

Taking into account Eq.(\ref{23}) the expansion (\ref{19}) for inverse 
surface correlation function can be written in the form 

\be
[G^{(0,2)}(p;m,c)]^{-1}=\kappa - m + Z_{\parallel}^{-1}(m+c) - 
[\sigma_{0}(p;m,c_0)-\sigma_{0}(0;m,c_0)].\label{25}
\ee

According to Eq.(\ref{25}), the surface enhancement renormalization of 
the inverse surface correlation function reduces to the subtraction from 
each diagram contribution the same diagram at 
zero external momentum. Thus, the uv divergences connected with 
the presence of the surface in the system are mutually canceled.

The iteration process of finding solution of the Eq. (\ref{23}) implies
that at zero-order we put $\delta c=0$ and find the first-order solution 
\be
\delta c^{(1)} = (Z_{\parallel}^{-1}-1)(m+c) + 
\sigma_{1}(0;m,c_0=c).\label{26}
\ee
Taking into account that we have dealt with the special transition where 
$c=0$, for the renormalization factor $Z_{\parallel}$ according to 
Eq.(\ref{12}) at one-loop order we obtain
\be
(Z_{\parallel}^{-1})^{(1)}=-\left.2 m \frac{\partial 
\sigma_{1}(p;m,0)}{\partial 
p^2}\right|_{p=0}=-\frac{\bar{t}_{1}^{(0)}}{4},\label{27}
\ee
where $\bar{t}_{1}=t_{1}/(8 \pi m)$. At two-loop order, the first-order 
contribution $\delta c^{(1)}$ must be taken into account in 
$\sigma_{1}$ because in common we have 
\be
\sigma_{1}(p;m,c_{0}) = 
\left.\sigma_{1}(p;m,c_{0})\right|_{0}+\left.\frac{\partial 
\sigma_{1}(p;m,c_{0})}{\partial c_{0}}\right|_{0}\delta 
c^{(1)}+...\label{28} 
\ee
The uv singular part of the second term of this expansion remove the 
divergence of the correspondent surface subgraph in $\sigma_{4}$, denoted 
via "+". As we can see from (\ref{19}), the similar subgraphs arise twice 
in $\sigma_{3}$. But the contribution $\sigma_{3}$ is uv finite because 
the Dirichlet propagator $G_{D}$ vanishes whenever $z$ or $z^{'}$
approach zero. 
Thus, the perturbation expansion of the inverse surface correlation 
function of our system up to two-loop order after bulk and surface 
enhancement renormalization can be written in the form 

\begin{eqnarray} 
&& [G^{(0,2)}]^{-1}=\kappa + (Z_{\parallel}^{-1})^{(1)} m - 
(\sigma_{1}(p;m,0)-\sigma_{1}(0;m,0)) +\left.\frac{\partial\sigma_{1} 
(0;m,c_{0})} {\partial c_{0}}\right|_{0} \delta c^{(1)}\nonumber\\ 
&& - \left.\frac{\partial\sigma_{1} (p;m,c_{0})} {\partial c_{0}}\right|_{0} 
\delta c^{(1)} - \sum_{i=2}^{4}\sigma_{i}(p;m,0).\label{29} 
\end{eqnarray} 

Following to the definition (\ref{12}) we obtain the next expansion for 
the renormalization factor $Z_{\parallel}$ up-to to loop-order  
\be
Z_{\parallel}^{-1} = 
1-\frac{\bar{t}_1^{(0)}}{4}+\frac{\bar{t}_1^{(0)2}}{4}(\frac{1}{2}-ln 
2)-\left.\lim_{p\to 0}\frac{m}{p}\frac{\partial}{\partial p} 
\frac{\partial\sigma_1(p;m,c_0)}{\partial c_0}\right|_{0} 
\sigma_{1}(0;m,0) - \lim_{p\to 0}\frac{m}{p}\frac{\partial}{\partial p} 
\sum_{i=2}^{4}\sigma_{i}(p;m,0).\label{30} \ee
 
 Performing the integration of Feynman integrals in (\ref{30}) by analogy 
 with \cite{DSh98} we derive the result 
\be
Z_{\parallel}^{-1}(\bar u_0,\bar v_0)=1-{\bar t_1^{(0)}\over 4} 
+\frac{\bar t_1^{(0)2}}{4}(\frac{1}{2}-ln 2 + ln^{2} 2) + \bar t_2^{(0)} 
A,\label{31} 
\ee
where the constant $A$ arise from the two-loop contribution in
(\ref{30}), 

\be
A\simeq{0.202428}.\label{32}
\ee
 
Here the renormalization factor $Z_{\parallel}$ is expressed as a 
second-order series expansion in powers of {\it bare} dimensionless 
parameters 
$\bar u_0=u_0/(8\pi m)$ and $\bar v_0=v_0/(8\pi m)$.
The corresponding weighting factors $\bar t_1^{(0)}$ and $\bar t_2^{(0)}$
are obtained by replacements
$(u_0,v_0)\to (\bar u_0,\bar v_0)$ in the original combinations
$t_1^{(0)}$ and $t_2^{(0)}$ from (\ref{ti}). As it is usual in {\it 
super}renormalizable theories the next step is the vertex renormalizations 

\begin{mathletters}
\begin{eqnarray}
\bar{u}_{0}&=&\bar{u}\Big(1+\frac{n+8}{6}\,\bar{u}+\bar{v}\Big),\nonumber\\
\bar{v}_{0}&=&\bar{v}\Big(1+\frac{3}{2}\,\bar{v}+2\, \bar{u}\Big)\,.\label{33}
\end{eqnarray}
\end{mathletters}

Again, the vertex renormalization at $d=3$ is a finite reparameterization.
The result for a modified series expansion of $Z_{\parallel}$ in terms 
of a new renormalized coupling constants $\bar u$ and $\bar v$, 
normalized in a standard fashion (\ref{33}), is

\begin{eqnarray}
Z_{\parallel}^{-1}(\bar u, \bar v)&=&
1-{n{+}2\over 12}\bar u - {\bar v\over 4}
+{n{+}2\over 3}\left(A-\frac{1}{4}+{n{+}2\over 12}(ln^{2} 
2-ln 2)\right)\bar u^2\nonumber\\
&+&\left(A-\frac{1}{4}+{1\over 4}(ln^{2} 2-ln 2)\right)\bar 
v^2 + 2\left(A-\frac{1}{4}+{n{+}2\over 12}(ln^{2} 2-ln 2)\right)\bar 
u\bar v.\label{34} 
\end{eqnarray}

Combining the renormalization factor $Z_{\parallel}(\bar u, \bar v)$ 
together with the one-loop pieces of the beta functions
\begin{mathletters}
\begin{eqnarray}
\beta_{\bar u}(\bar u, \bar v)&=&-\bar u\Big(1-{n{+}8\over 6}\;\bar u-\bar 
v\Big),\nonumber\\
\beta_{\bar v}(\bar u, \bar v)&=&-\bar v\Big(1-{3\over 2}\,\bar 
v-2\bar u\Big),\label{35} 
\end{eqnarray}
\end{mathletters}
through (\ref{16}), we obtain the desired series expansion for 
$\eta_{\parallel}^{sp}$

\begin{eqnarray}\label{etafin}
&&\eta_{\parallel}^{sp}(u,v)=-{n{+}2\over {2 (n{+}8)}}\,u-{v\over 6}\\
&&+12{(n{+}2)\over (n{+}8)^2}A(n) u^2+{4\over 9} 
A(1) v^2 + {8\over n{+}8}A(n) uv ,\nonumber
\end{eqnarray}
where $A(n)$ is a function of the replica number $n$, defined as 
\begin{equation}
A(n)=2A+{n{-}10\over 48}+\frac{n+2}{6}(ln^{2} 2 - ln 2),\label{36}
\end{equation}
and renormalized coupling constants $u$ and $v$, normalized in a 
standard fashion $u{=}{n+8\over 6}{\bar{u}}$ and $v{=}{3\over 
2}{\bar{v}}$.

In fact, the last expression (\ref{etafin}) for $\eta_{\parallel}^{sp}$ 
provides a result for the {\it cubic anisotropic} model given by the effective 
Hamiltonian (\ref{6}) with general number $n$ of order-parameter 
components. In the case of infinite space, this cubic anisotropic model 
attracted much attention very recently (see e.g. \cite{PS00,Var00,CPV00} 
and references therein).

In the present paper we restrict our discussion to the case of {\it 
semi-infinite random Ising-like} systems by taking the replica limit $n\to 
0$. According to (\ref{etafin}) we obtain the next two-loop expansion for 
the surface critical exponent $\eta_{\parallel}^{sp}$ at the special 
transition  

\begin{equation}
\!\!\eta_\|^{sp} =-{u\over 8}-{v\over 6}+{3\over 8}A(0)u^2+
{4\over 9}A(1)v^2+A(0)uv.\label{37}
\end{equation}

As it is well known, the knowledge of one surface critical exponent gets 
access via usual scaling relations \cite{D86} to the other surface 
critical exponents. For convenience further below we suppress the 
superscript {\it sp} at the surface critical exponents.

\section{Calculation of the surface critical exponents}

The present section is devoted to numerical calculation of the surface 
critical exponents at the special transition. The individual RG series 
expansions for other surface critical exponents can be derived from 
(\ref{37}) through standard scaling relations \cite{D86} (with $d=3$)

\begin{eqnarray} 
&& \eta_{\perp} = \frac{\eta + 
\eta_{\parallel}}{2}, \nonumber\\ 
&& \beta_{1} = \frac{\nu}{2} 
(d-2+\eta_{\parallel}), \nonumber\\ 
&& \gamma_{11}=\nu(1-\eta_{\parallel}), \nonumber\\ 
&& \gamma_{1}= \nu(2-\eta_{\perp}), \label{sc}\\
&& \Delta_{1}= \frac{\nu}{2} (d-\eta_{\parallel}), \nonumber\\ 
&& \delta_{1} = \frac{\Delta}{\beta_{1}} = 
\frac{d+2-\eta}{d-2+\eta_{\parallel}}, \nonumber\\ 
&& \delta_{11} = \frac{\Delta_{1}}{\beta_{1}}= 
\frac{d-\eta_{\parallel}}{d-2+\eta_{\parallel}}\;.\nonumber
\end{eqnarray}

Each of these critical exponents characterizes certain properties of 
the system with the surface in the vicinity of the critical point (see 
\cite{UShH00}). The values $\nu$, $\eta$, and $\Delta=\nu(d+2-\eta)/2$ 
are the standard bulk exponents. The correspondent series expansions for 
$\nu$ and $\eta$ at $d=3$ are given by (\cite{Jug83,SSh81,Sh88}) 

\begin{eqnarray}
&& \nu=\frac{1}{2} \left[1+ \frac{v}{6} + \frac{(n+2)}{2(n+8)} u 
\right.\nonumber\\
&& \left.-\frac{1}{324} \left[ \frac{11}{9} v^{2}- \frac{2}{n+8}(27n-38) 
u v - \frac{3 (n+2)}{(n+8)^2} (27n-38) u^{2} \right]\right],\nonumber\\
&& \eta=\frac{8}{27} \left[ \frac{ v^{2}}{27} + 
\frac{2 u v}{3(n+8)} + \frac{(n+2)}{(n+8)^2} u^{2}\right]. \label{38}
\end{eqnarray}

For each of surface critical exponents we obtain from (\ref{sc}) and 
(\ref{etafin}) at $d=3$ a double series expansion in powers of $u$ and $v$ 
truncated at the second order
\be
f(u,v)=\sum_{j,l\geq o} f_{jl} u^{j} v^{l}.\label{39}
\ee
Since perturbation expansions of this kind 
are generally divergent \cite{LO}, the sufficiently powerful resummation 
procedure of the series is essential to obtain accurate estimates of the 
critical exponents. One of the simplest way is to calculate for each 
quantity we consider a sequence of rational Pad\'e approximants in two 
variables from the original series expansions. This should work already 
well when the series behave in lowest orders `in a convergent fashion'. 
Besides, if the series are alternating in sign \cite{BNM78}, we can use  
more modern Pad\'e-Borel resummation procedure \cite{BNGM76} for their 
analysis.

The results of our Pad\'e and Pad\'e-Borel analysis for the surface 
critical exponents at the special transition are represented in Table I. We 
evaluate the exponents at the standard RG random fixed point {\cite{Jug83}}
\begin{equation}\label{NT}
u^{*}=-0.60509, \quad\quad\quad v^{*}=2.39631.
\end{equation}

The values $[0/0]$, $[1/0]$, and $[2/0]$ are simply the direct partial sums
up to the zeroth, first, and second orders, respectively. Pad\'e 
approximants $[0/1]$ and $[0/2]$ represent the partial sums of the 
inverse series expansions up to the first and second order.

Besides, by analogy with \cite{UShH00} we performed the 
calculation of nearly-diagonal two-variable rational approximants of the 
type 
\be
[11/1] = \frac{1 + a_{1} u + \bar{a}_{1} v + a_{11} uv}{1 + b_{1} u +
\bar{b}_{1} v }\label{40}
\ee
and
\be
[1/11] = \frac{1 + a_{1} u + \bar{a}_{1} v}{1 + b_{1} u +
\bar{b}_{1} v+ b_{11} uv },\label{41}
\ee
which are given in eighth and ninth columns of Table I. 

The second column of Table I contains the ratios  
of magnitudes of first- ($O_1i$) and second-order 
($O_2i$) perturbative corrections appearing in inverse 
series expansions of our critical exponents. The larger (absolute) 
values of this ratios correspond to the better apparent convergence of 
truncated series. It is easy to see that the series 
of inverse expansions for all critical exponents, except $\beta_{1}$, are 
alternating in sign and consequently adapted to the above-mentioned 
Pad\'e-Borel resummation analysis (see Appendix 1). Among the direct series 
the situation is more complicated. The ratios of first- $O_{1}$ and 
second-order $O_{2}$  perturbative corrections of the direct series 
expansions for the critical exponents $\delta_{1}$, $\gamma_{11}$ and 
$\gamma_{1}$ are positive \cite{mit3}. This means that the signs of the 
first- and second-order corrections do not alternate and hence the 
correspondent series are not suitable to the Pad\'e-Borel resummation 
technique, since the $[11/1]$ approximant of the Borel transform have a 
pole in the integration range. But these series are 
slowly convergent, because the contribution of the second-order are 
considerably less than contribution of the first-order. For example the 
ratio $O_{1}/O_{2}$ for the critical exponent $\Delta_{1}$ will be equal 
35.1. Thus the above mentioned series adapted to the 
Pad\'e analysis. It should be noted, that a very similar situation has been 
meet in the analysis of the surface critical exponents perturbation series 
expansions at the ordinary transition in pure \cite{DSh98} and  
quenched dilute semi-infinite Ising-like systems \cite{UShH00}.

The results of Pad\'e-Borel analysis of the inverse 
series expansions are given in the last column of the Table I. These 
values give the most reliable numerical estimates. In common, we assume 
that the results obtained from the Pad\'e-Borel analysis $R^{-1}$ are the 
best we could achieve from the available knowledge about the series 
expansions in the frames of the present approximation scheme 
\begin{eqnarray}
&&\eta_{\parallel}=-0.238,\quad \Delta_{1}=1.101,\quad 
\eta_{\perp}=-0.116,\quad \beta_{1}=0.258,\nonumber\\
&&\gamma_{11}=0.845, \quad \gamma_{1}=1.442, \quad \delta_{1}=6.343, 
\quad \delta_{11}=4.172.\label{42}
\end{eqnarray}
Here the estimate of $\beta_{1}$, for which no Pad\'e-Borel approximant 
$R^{-1}$ exists, has been derived from the scaling relation 
$\beta_{1}=\frac{\nu}{2}(d-2+\eta_{\parallel})$, where $\nu=0.678$ 
\cite{Jug83} and the above value of $\eta_{\parallel}=-0.238$.
 
The deviations of estimates (\ref{42}) from the other second-order 
estimates of the table might serve as a rough measure of the achieved 
numerical accuracy.

\section{$\sqrt{\epsilon}$ expansion} 

As was mentioned above, there is an alternative method to analyze the 
influence of randomness on the critical behavior based
on the renormalization-group approach introduced by Harris and Lubensky 
\cite{HL74}. This method was employed by Ohno and Okabe \cite{OO92} for 
study critical behavior of semi-infinite systems with a Gaussian 
randomness in $4-\epsilon$ dimensions in the frames of $\sqrt{\epsilon}$ 
expansion for obtaining the two-loop approximation for correlation 
function and deriving corresponding series expansions for the surface 
critical exponents $\eta_{\parallel}$ and $\eta_{\perp}$.  
Their results in the case of special transition at $n=1$ (see 
\cite{OO92}) with corresponding changes of coupling 
constants normalizations ($u\to v/24$, $w\to -u/3$) in accordance with our 
notations are written in the form 
\begin{eqnarray}
&&\eta_\|=-{u\over 3} - {v\over 2} + {7\over 12}\,u^2 + {11\over 
12}\,v^2+{7\over 4}\,u\,v + O(\epsilon^{3\over 2})\nonumber\\
&&\eta_\perp=-{u\over 6}-{v\over 4}+{11\over 36}\,u^2+{23\over 
48}\,v^2+{11\over 12}\,u\,v + O(\epsilon^{3\over 2}).\label{43}
\end{eqnarray} 

Unfortunately, the surface critical exponents $\eta_{\parallel}$ and 
$\eta_{\perp}$ have been obtained only to the first order in 
$\sqrt{\epsilon}$ from these equations in \cite{OO92}. In the present 
paper we derive the next term in $\sqrt{\epsilon}$ expansion for above 
mentioned surface critical exponents using the fixed-point values up 
to $O(\epsilon)$ \cite{Sh77,JK77}

\begin{eqnarray} 
&&u^*=-3 \sqrt{\frac{6 \epsilon}{53}}+18 
\frac{110+63\zeta(3)}{53^2}\epsilon,\nonumber\\ 
&&v^*=4\sqrt{\frac{6 
\epsilon}{53}}-72\frac{19+21 \zeta(3)}{53^2}\epsilon,\label{fixpe} 
\end{eqnarray}
where $\zeta (3)\simeq 1.2020569$ is the Riemann $\zeta$-function, and the usual
geometric factor $K_d=2^{1-d}\pi^{-d/2}/\Gamma(d/2)$ has been absorbed into 
the redefinitions of the coupling constants. As a result we obtain 

\begin{eqnarray}
&& \eta_{\parallel}=-\sqrt{\frac{6 \epsilon}{53}}+ 
\frac{756\zeta(3)-641}{2\cdot 53^2} \epsilon,\nonumber\\
&& \eta_{\perp}=-\sqrt{\frac{3 \epsilon}{106}} + 
\frac{378 \zeta(3)-347}{2\cdot 53^2}\epsilon.\label{44}
\end{eqnarray}

Taking into account scaling relations for surface critical 
exponents and $\sqrt{\epsilon}$ expansions for random bulk exponents $\nu$ 
and $\eta$ \cite{Sh77,JK77} we obtain respective perturbation series 
expansions for other surface critical exponents 
\begin{eqnarray}
&& 
\Delta_{1}=\frac{3}{4}+\frac{5}{8}\sqrt{\frac{6\epsilon}{53}}+
\frac{3523-3780\zeta(3)}{16{\cdot}53^2}\epsilon,\nonumber\\ 
&& \beta_{1}=\frac{1}{4}-\frac{1}{8}\sqrt{\frac{6 
\epsilon}{53}}+{3\over 16}\frac{(252 
\zeta(3)-461)}{53^2}\epsilon,\nonumber\\ 
&& \gamma_{11}=\frac{1}{2}+\frac{3}{4} \sqrt{\frac{6 \epsilon}{53}}-
\frac{2268\zeta(3)-2453}{8{\cdot}53^2}\epsilon,\nonumber\\ 
&& \gamma_{1}=1 + \frac{3}{4}\sqrt{\frac{6 \epsilon}{53}} - 
\frac{3}{4}\frac{(378 \zeta(3)-347)}{53^2}\epsilon,\nonumber\\ 
&& \delta_{1}=5+5\sqrt{\frac{6 \epsilon}{53}} - 
3 \frac{(630 \zeta(3)-1073)}{53^2}\epsilon,\nonumber\\
&& \delta_{11}=3+4\sqrt{\frac{6 \epsilon}{53}}-
2 \frac{(756 \zeta(3)-1277)}{53^2}\epsilon.\label{45} 
\end{eqnarray}

Similarly as in the case $d=3$ of the previous section,
we perform a Pad\' e analysis of our $\sqrt{\epsilon}$ 
expansions at $\epsilon =1$. The numerical values of surface critical 
exponents obtained in this way are represented in Table II.
It should be noticed that as was shown in \cite{FHY99,FHY00}, the 
$\sqrt{\epsilon}$ expansion is not Borel summable. 

The Pad\'e approximants [1/0] for the 
exponents $\eta_{\parallel}$ and $\eta_{\perp}$ reproduce the first-order 
results obtained by Ohno and Okabe \cite{OO92}. On the other hand, the 
other exponents, $\beta_{1}$, $\gamma_{11}$ and $\gamma_{1}$ slightly 
differ from the previous results \cite{OO92}
\be
\beta_{1}=0.17, \quad\quad \gamma_{11}=0.78, \quad\quad 
\gamma_{1}=1.26.\label{46}
\ee
The reason is that we calculated our $[1/0]$ estimates directly from each 
respective $\sqrt{\epsilon}$ expansion, while in Ref. \cite{OO92} they 
were obtained from the scaling relations using the above mentioned 
numerical vales of $\eta_{\parallel}$ and $\eta_{\perp}$ (\ref{44}). In 
addition we performed the analysis of the correspondent series expansions 
for the surface magnetic shift exponent $\Delta_{1}$, exponents 
$\delta_{1}$ and $\delta_{11}$ which give relations between the surface 
magnetization and the surface and bulk external magnetic fields, 
respectively.

It can be easily verified that the above first-order approximants 
of the critical exponents satisfy the scaling relation 
\be
\eta_{\perp} = (\eta + \eta_\parallel)/2 \label{47}
\ee 
with the value of the bulk exponent 
$\eta=-{\epsilon\over 106}+O(\epsilon^{3\over 2})$  \cite{L75,X75,GL76}.

Comparing the results given in Tables I and II we see that
the values of the first-order approximants, denoted by [1/0] 
and [0/1] in both cases, are of comparable magnitudes.
But, on the other hand, the values of the second-order approximants
are significantly different in both tables. As was shown previously 
\cite{B87,McK94,ShAnSo97,FHY99,FHY00}, $\sqrt{\epsilon}$ series expansions 
possess rather irregular structure and are practically unsuitable for 
subsequent resummation and are ineffective for obtaining reliable numerical 
estimates. Our results confirm this assumption. If we try to reproduce the 
numerical value of $\eta$ (see (\ref{47})) from our second-order data of 
Table II, we always obtain negative values. But, this does not agree with 
the sufficiently precise positive results of massive field-theoretic 
approach for random bulk systems in three dimensions up to two-loop 
\cite{Jug83,HSh92}, three-loop \cite{MS84,Sh88,FHY99} and to the 
four-loop \cite{MSSh89} order. Besides, very 
recently was obtained the value $\eta=0.025\pm 0.01$ in the frames of 
five-loop renormalization-group expansions \cite{PS00} and up-to six-loop 
order $\eta=0.030(3)$ \cite{PV00}). This discrepancy is not present in our 
calculation performed directly at $d=3$ (see Table I). From the surface 
scaling relation (\ref{47}) and the second-order results of Table I we 
obtain the value $\eta=0.031$, which quite well agree with previous 
estimates.

\section{Summary}

The main aim of the present article was an investigation of the influence 
of quenched bulk randomness on the surface critical behavior of 
semi-infinite Ising-like systems at the special transition. To summarize, 
we have calculated the surface critical exponents of the special transition 
for such systems using two alternative technique: the massive field 
qtheory approach directly at $d=3$ dimensions up-to two-loop order, and the 
$\sqrt{\epsilon}$ expansion at $d=4-\epsilon$ dimensions to the order of 
$O((\sqrt{\epsilon})^2)$. In the last case we extend up to the next to 
leading order, the previous first-order results obtained by Ohno and Okabe 
\cite{OO92}. But, the $\sqrt{\epsilon}$ expansions possess rather irregular 
structure, as was shown in \cite{FHY99,ShAnSo97,FHY00}. This makes them 
practically unsuitable for subsequent resummation and ineffective for 
getting quantitative numerical estimates. 

In both cases the resummation of obtained perturbation series 
expansions for surface critical exponents was performed using Pad\'e 
analysis. Moreover, in the case of the massive field theory 
the resulting two-loop series expansions were resummed by means of 
more precise Pad\'e-Borel resummation technique. In the previous section we 
have discussed some merits of using the massive field-theoretic approach 
directly in $d=3$ dimensions for obtaining most accurate numerical 
estimates for critical exponents. Thus, the best estimates for surface 
critical exponents of semi-infinite systems with quenched bulk disorder at 
the special transition, which we can obtain in the frames of the present 
approximation scheme, are

\begin{eqnarray}
&&\eta_{\parallel}=-0.238,\quad \Delta_{1}=1.101,\quad 
\eta_{\perp}=-0.116,\quad \beta_{1}=0.258,\nonumber\\
&&\gamma_{11}=0.845, \quad \gamma_{1}=1.442, \quad \delta_{1}=6.343, 
\quad \delta_{11}=4.172.\label{48}
\end{eqnarray}

The obtained results are evidently different from results obtained for pure
semi-infinite Ising-like systems \cite{DSh98,DSh94}

\begin{eqnarray}
&&\eta_{\parallel}=-0.165,\quad \Delta_{1}=0.997,\quad 
\eta_{\perp}=-0.067,\quad \beta_{1}=0.263,\nonumber\\
&&\gamma_{11}=0.734, \quad \gamma_{1}=1.302, \quad \delta_{1}=5.951, 
\quad \delta_{11}=3.791\label{48}
\end{eqnarray}

and confirm the assumption that the presence of quenched bulk  
disorder affects the critical behavior of boundaring surface. So, in the 
case of special transition, similarly as in the case of previously 
investigated ordinary transition \cite{UShH00}, the new set of surface 
critical exponents appear. It should be mentioned that at the present time 
the value of the surface crossover exponent $\Phi$ for semi-infinite 
systems with quenched bulk disorder is still an open question. This 
question will be the topic of our next publication.

\section*{Acknowledgments}
We should like to express our warmest thanks Dr. M. Shpot for a useful 
discussion and Professor Y. Okabe for reading the manuscript. This work was 
supported by the National Science Council of the Republic of China (Taiwan) 
under Grant No. NSC 89-2112-M-001-084. 

\section*{Appendix}

As was mentioned above, the power series expansions of surface 
critical exponents (see Eq.(\ref{39})) are in general not convergent. In 
order to obtain meaningful and rather accurate numerical estimates  
we must apply to them sufficiently powerful "resummation" procedure.
In the present paper we employ a two-variable resummation technique 
\cite{Jug83,MS84,Sh88,MSSh89,HSh92} which is a simple generalization of 
the single-variable Pad\'e-Borel method. The starting point of this 
calculation is construction for truncated power series (\ref{39}) the 
Borel transform 
\be
F(ut, vt)=\sum_{j,l\geq o} \frac{f_{jl}}{(j+l)!} (ut)^{j} 
(vt)^{l}.\label{A1} 
\ee
Then we construct the rational approximant $F^{B}(x,y)$ 
\be
F^{B}(u,v)= \frac{1+a_{10}u+a_{01}v+a_{11}uv}{1+b_{10}u+b_{01}v},\label{A2}
\ee
which is extrapolation of the Borel transform (\ref{A1}). It is clear 
that at $u=0$ or $v=0$ we obtain from (\ref{A2}) the usual [1/1] Pade 
approximant. The coefficients $a_{jl}$ and $b_{jl}$ are expressed via the 
appropriate expansion coefficients of the initial function $f(u,v)$ 
(\ref{39}) \begin{eqnarray} &&a_{01}=g_{10}+b_{10}, \quad\quad 
b_{10}=-g_{20}/g_{10},\nonumber\\ &&a_{01}=g_{01}+b_{01}, \quad\quad 
b_{01}=-g_{02}/g_{01},\nonumber\\ 
&&a_{11}=g_{11}+b_{10}g_{01}+b_{01}g_{10}, \label{A3} 
\end{eqnarray}
where $ g_{jl}=f_{jl}/(j+l)!$. Hence, for the resummed function by means 
of the Pad\'e-Borel resummation technique we obtain  
\be 
{\bar{f}}(u,v)= \int_{0}^{\infty} F^{B}(ut, vt) e^{-t} 
dt.\label{A4} 
\ee
This technique has been applied to the direct power series expansions of 
surface critical exponents to find the  corresponding Pad\'e-Borel 
approximants $R^{-1}$.

\begin{table}[htb]
\caption{Surface critical exponents of the special transition for $d=3$ 
up to two-loop order at the random-fixed point $
u^{*}=-0.60509,  v^{*}=2.39631$.}
\label{tab2}
\begin{center}
\begin{tabular}{rrrrrrrrrr}
\hline
$ exp $~&~$\frac{O_{1i}}{O_{2i}}$~&~$[0/0]$~&~
$[1/0]$~&~$[0/1]$~&~$[2/0]$~&~$[0/2]$~&~$[11/1]$~&~$[1/11]$~&~$ 
R^{-1} $ \\ 
\hline
$\eta_{\parallel}$ & -23.82 & 0.00 & -0.324 & -0.245 & -0.205 & 
-0.237 & -0.244 & -0.238 & -0.238 \\

$\Delta_{1}$ & -3.39 & 0.75 & 1.074 & 1.229 & 1.083 & 1.046 & 1.083
 & 1.090 & 1.101 \\

$\eta_{\perp}$ & -3.35 & 0.00 & -0.162 & -0.139 & -0.087 & -0.102 & 
-0.115 & -0.114 & -0.116\\

$\beta_{1}$ & 0.00 & 0.25 & 0.25 & 0.25 & 0.263 & 0.263 & 
--- & --- & ---\\

$\gamma_{11}$ & -3.14 & 0.50 & 0.824 & 0.979 & 0.825 & 0.783 
& 0.825 & 0.834 & 0.845\\

$\gamma_{1}$ & -2.56 & 1.00 & 1.405 & 1.680 & 1.410 & 
1.327 & 1.410 & 1.421 & 1.442\\

$\delta_{1}$ & -1.41 & 5.00 & 6.619 & 7.394 & 7.062 & 5.521 & 
6.205 & 6.236 & 6.343\\

$\delta_{11}$ & -1.40 & 3.00 & 4.295 & 5.279 & 3.926 & 3.418 & 
4.032 & 4.070 & 4.172\\
    
\end{tabular}
\end{center}
\end{table}

\begin{table}[htb]
\caption{Surface critical exponents of the special transition from the 
$\sqrt{\epsilon}$  expansion. } 
\label{tab3}
\begin{center}
\begin{tabular}{rrrrrrr}
\hline
$ exp $~&~$[0/0]$~&~
$[1/0]$~&~$[0/1]$~&~$[2/0]$~&~$[0/2]$~&~$[1/1]$ 
\\ 
\hline
$\eta_{\parallel}$ & 0.00 & -0.336 & -0.252 & -0.289 & -0.287 
 & -0.295\\

$\Delta_{1}$ & 0.75 & 0.960 & 1.016 & 0.938 & 0.917 
 & 0.940\\

$\eta_{\perp}$ & 0.00 & -0.168 & -0.144 & -0.149 & -0.151 & 
-0.151\\

$\beta_{1}$ & 0.25 & 0.208 & 0.210 & 0.197 & 0.198 & 
0.194\\

$\gamma_{11}$ & 0.50 & 0.752 & 0.838 & 0.740 & 0.714 & 
0.741\\

$\gamma_{1}$ & 1.00 & 1.252 & 1.338 & 1.224 & 
1.190 & 1.227\\

$\delta_{1}$ & 5.00 & 6.682 & 7.535 & 7.019 & 6.729 & 7.104\\
	
$\delta_{11}$ & 3.00 & 4.346 & 5.441 & 4.608 & 4.232 & 
4.671\\	 
\end{tabular}
\end{center}
\end{table}


\begin{thebibliography}{10}
\bibitem{mit2} The corresponding phase diagram of the semi-infinite systems 
was represented in \cite{D86,D97}.
\bibitem{LO}
For this kind of series, the expansion coefficients at large orders of 
perturbation theory grow nearly factorial. In fact, this 
is an intuitive picture conveyed from the theory of bulk regular
  systems. Much less is known about the large-order behavior of perturbative
  expansions pertaining to infinite random systems (see Refs.
  \cite{lo87,McK94,AMR00}), especially at large space dimensionalities. To our
  knowledge, there are no explicit results on large orders for surface
  quantities, even in the absence of disorder.
\bibitem{mit3} For the simplicity we do not present the ratio 
$O_{1}/O_{2}$ here.
\end{thebibliography}
\end{document}